\documentclass[twoside,final]{article}

\usepackage{natbib}
\usepackage[dvips]{epsfig}
\usepackage{amsfonts}
\usepackage{amssymb}
\usepackage[mathscr]{eucal}
\usepackage{latexsym,fullpage,picinpar}

\newcommand{\N}{\mathbb{N}}
\newcommand{\Z}{\mathbb{Z}}

\newcommand{\sub}{\underline}

\newtheorem{definition}{Definition}

\newtheorem{theorem}{Theorem}
\newtheorem{example}{Example}

\newtheorem{lemma}{Lemma}
\newtheorem{corollary}{Corollary}

\newenvironment{proofUsing}[1]
	{\begin{trivlist}\item[]{\em Proof #1.}}
	{\hspace*{\fill}$\square$\end{trivlist}}

\newenvironment{proof}
	{\begin{trivlist}\item[]{\em Proof.}}
	{\hspace*{\fill}$\square$\end{trivlist}}

\begin{document}

\title{Tiling with bars under tomographic constraints}

\footnotetext[1]{
	Universit\'e de Paris-Sud, 
	LRI b\^at 490, 
	91405 Orsay, 
	France,
\texttt{durr@lri.fr}.
	partially supported by the Programme de Coop\'{e}ration
	Franco-Chilienne du Minist\`{e}re des affaires Etrang\`{e}res
	(France). Corresponding author.  Work was done while the author 
	was affiliated with
	\emph{Computer Science Institute,
	The Hebrew University of Jerusalem,
	Givat Ram, Jerusalem, Israel.}}

\footnotetext[2]{
	DIM, Universidad de Chile,
	casilla 170-3 correo 3,
	Santiago, Chile,
\texttt{egoles@dim.uchile.cl}.}

\footnotetext[3]{
	Partially supported by project Fondap on Applied Mathematics.}

\footnotetext[5]{
	Partially supported by project Fondecyt Nr.1990616 and ECOS.}

\footnotetext[4]{
	CMM (CNRS UMR 2071) and DIM, Universidad de Chile,
	\texttt{irapapor@dim.uchile.cl}.}

\footnotetext[6]{
	Grima, IUT Roanne,
	42334 Roanne cedex, France \emph{or}
	LIP, ENS-Lyon, CNRS umr 8512,
	69364 Lyon cedex 07, France,
\texttt{\small Eric.Remila@ens-lyon.fr}.}

\author{
	Christoph D\"{u}rr${}^{1}$
\and
	Eric Goles${}^{2,3}$
\and
	Ivan Rapaport${}^{3,4,5}$
\and
	Eric R\'{e}mila${}^{3,6}$
}

\date{}

\maketitle

\begin{abstract}
We wish to tile a rectangle or a torus with only vertical and
horizontal bars of a given length, such that the number of bars in
every column and row equals given numbers. We present results for
particular instances and for a more general problem, while leaving open 
the initial problem.
\end{abstract}

\paragraph{Keywords:}
Discrete Tomography, Domino Tiling

\section{Introduction}

In general terms, \emph{tomography} is the area of reconstructing
objects from lower dimensional projections. We consider the problem of
reconstructing a rectangular grid from projections on the columns and
on the rows. Think of the grid as a layer in a crystal, and in order
to measure it, we send beams through the crystal from two orthogonal
directions.  Measurements will give us quantitative information about
columns or rows of the grid~\cite{Kal95}.  Consider the problem, where
each cell of the grid (think of it as an atom) is matched with at most
one of its immediate neighbors (think of it as a chemical connection).
Physicists call them \emph{monomer-dimer systems}.  Many research has
been done about counting the number of configurations of such a
system~\cite{KRS96}, or about almost uniform randomly sampling
configurations~\cite{LRS01}.

\begin{figwindow}[3,r,\epsfig{file=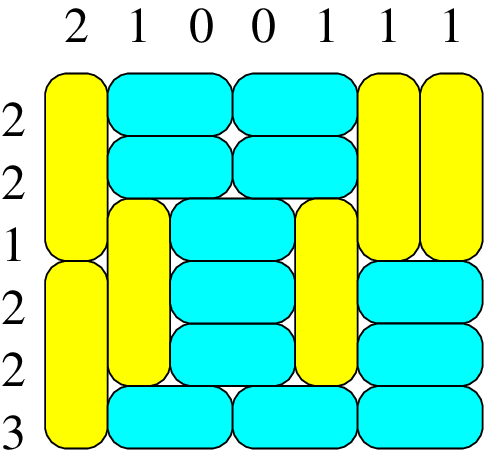,width=35mm},%
{A tiling and its projections.}]  We are interested in the
particular problem, were each cell is matched to exactly one
neighboring cell.  These objects correspond to domino tilings of the
grid.  A measurement will reveal the number of vertical dominoes in
each column and the number of horizontal dominoes in each row.  Given
these numbers we wish to reconstruct the grid, or any grid which
satisfies the projection constraints.  As a natural generalization of
this combinatorial problem we are interested in the tiling of the
grid with horizontal bars of length $h$ and vertical bars of length
$v$, for some integers $h,v$.  We call it the \textsc{Tiling with Bars
Reconstruction} problem.  Given a pair of column and row vectors
$(\sub m, \sub n)$ (\emph{tomographic constraints}) and integers $h,v$
we want to construct \emph{a} tiling with bars, such that $\sub m$
counts the number of vertical bars in the columns and $\sub n$ counts
the number of horizontal bars in the rows.  This problem has two
variants, whether we tile a rectangle or a torus.

The problem is left open by this paper, but we were able to find solutions
for subproblems and for a more general problem.  We summarize our results
in this table:
\end{figwindow}
\begin{center}
\begin{tabular}{l|cc}
				&rectangle		&torus \\ \hline 
tiling of a given sub-grid 	& NP-hard		&NP-hard\\
 general problem 		&open			&open\\
 $ \sub m $ is a uniform vector&quadratic algorithm	&open\\
 $ \sub n $ is also a uniform vector&
				algebraic characterization&
					algebraic characterization
\end{tabular}
\end{center}

The quadratic algorithm has been found independently by
Picouleau~\cite{Pic01} for a more general condition (see end of
section~\ref{sec-application}).

\section{Definitions}

\begin{figure}[htb]
\centerline{\epsfig{file=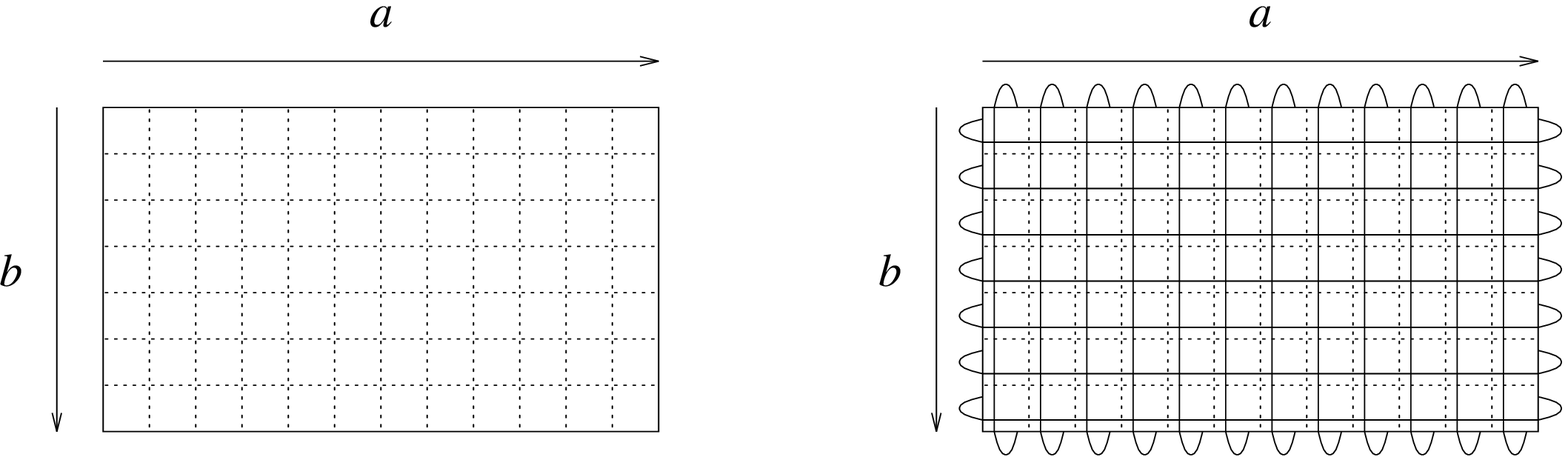,height=4cm}}
\caption{A rectangle \protect$R_{a\times b}\protect$ and a torus 
\protect$T_{a\times b}\protect$.}
					\label{fig:toro-rect}
\end{figure}

Let $ a,b\geq 1$. The rectangle $R_{a\times b}$ is the product
$\{0,\cdots ,a-1\}\times \{0,\cdots ,b-1\}$ and the torus $T_{a\times
b}$ is the product $\Z_a \times \Z_b$.  Columns are numbered from left
to right and rows from top to bottom (see Figure~\ref{fig:toro-rect}).
Each element of a rectangle or a torus is called a \emph{cell}.

Let $ h,v \geq 1$. The horizontal bar of length $h$ is the
rectangle $ R_{h\times 1} $ and the vertical bar of length $ v $ is
the rectangle $ R_{1\times v} $.  If the length is $ 2 $ we call the
bar a \emph{domino}.

A rectangle $R_{a\times b}$ (respectively a torus $T_{a\times b}$) is
said to be \emph{tillable} with the vertical and horizontal bars (of
lengths $ v $ and $ h $ respectively) if it can be partitioned into
those bars. The projections of such a tiling is the pair of vectors $
(\sub m,\sub n)=(m_{1}\cdots m_{a},n_{1}\cdots n_{b})\in \N ^{a}\times
\N ^{b} $ such that for every column $ i $, $ m_{i} $ is the number of
vertical bars in it, and for every row $ j $, $ n_{j} $ is the number
of horizontal bars in it.

We define the following reconstruction problems.

\begin{quote}
\noindent 
\textsc{Tiling a Rectangle (\textrm{respectively} Torus) 
	with Bars under Tomographic Constraints}
\begin{description}
\item [input]
	$ (\sub {m},\sub {n})\in \N ^{a}\times \N ^{b} $ and $h,v\geq 1$.
\item [output]
	a tiling of the $R_{a\times b}$ (respectively $T_{a\times b}$)
	with projections $ (\sub {m},\sub {n}) $.
\end{description}
\end{quote}

\section{Uniform constraints}

In this section we characterize valid instances for the special case
when the constraints vectors are uniform, that is $ \forall
i:m_{i}=m,\forall j:n_{j}=n $ for some integers $ m,n $. Both the
torus and the rectangle case are studied.

\subsection{The torus case}

The number of cells covered with vertical bars is $ amv $ and the
number of cells tiled with horizontal bars $ bnh $. Clearly these
numbers must add up to the total number of cells $ ab $. In this
section we show that this condition is sufficient for a torus tiling
to exist.

\begin{lemma}					\label{lemm:algebra}
If $a,b,h,v,m,n \geq 1$ are such that $ab=amv+bnh$ then there exist
$p,q,a',b' \geq 1$ satisfying
\begin{itemize}
\item $(p+q)=\gcd(a,b)$ and $a=(p+q)a'$ and $b=(p+q)b'$.
\item $nh=pa'$ and $mv=qb'$. 
\end{itemize}
\end{lemma}

\begin{proof}
If we denote $c={\gcd(a,b)}$, $a=ca'$, and $b=cb'$, then the equality
$ab = amv+bnh$ can be rewritten as $ca'b'=a'mv+b'nh$. It follows that
$(ca'- nh)b'=a'mv$. From Gauss theorem, $a' | (ca' - nh)$, and
therefore $a' | nh$.  In other words, there exists $p$ such
that $pa'=nh$.  By symmetry, there exists $q$ such that
$qb'=mv$.  Finally, notice that
\[
		p+q=\frac{c(cb'nh+ca'mv)}{c^{2}a'b'}=c.
\]
\end{proof}

\begin{theorem}				\label{prop:unif-torus}
Let $a,b,h,v,m,n \geq 1$.  Let $(\sub m,\sub n) = (m\cdots m,n\cdots
n) \in \N^{a} \times \N^{b}$.  The torus $T_{a \times b}$ is
$(\sub m,\sub n)$-tillable with the bars $R_{h \times 1}$ and $R_{1
\times v}$ if and only if $ab = amv + bnh$.
\end{theorem}

\begin{figure}[htb]
\centerline{\epsfig{file=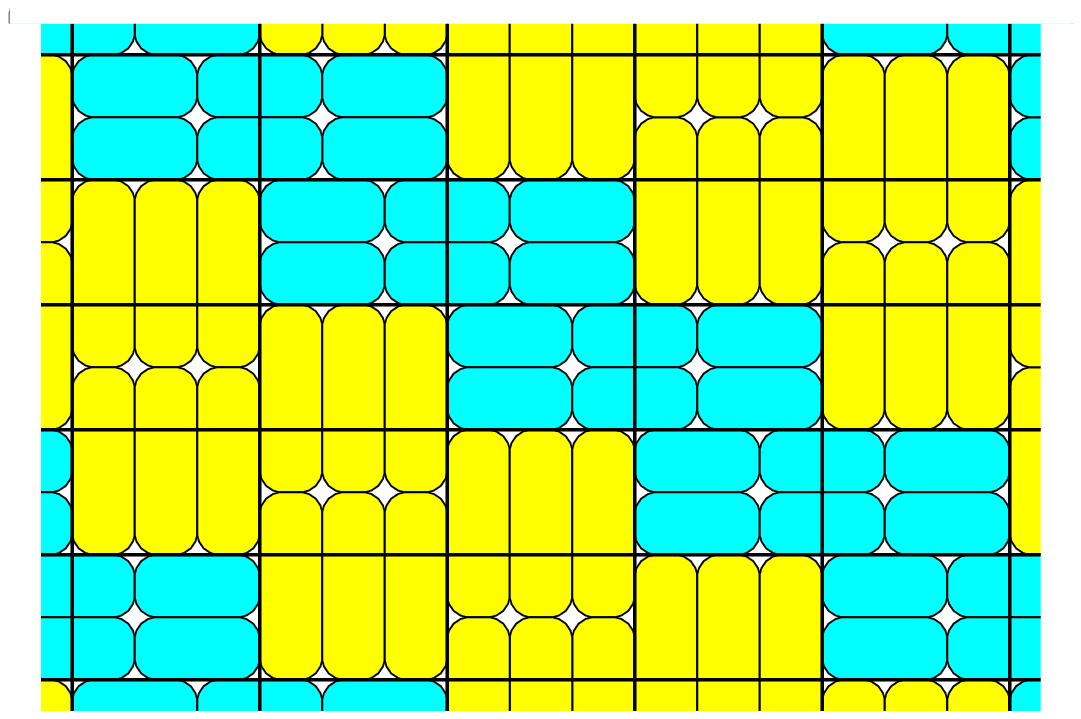,height=6cm}}
\caption{A \protect$ (2\cdots 2,3\cdots 3)\protect $-tiling of 
	\protect$ T_{15\times 10}\protect $ by \protect$ R_{2\times
	1}\protect $ and \protect$ R_{1\times 3}\protect $.}
						\label{fig:toro-unif}
\end{figure}

\begin{proof}
If we assume that the torus $T_{a \times b}$ admits an $(\sub m,\sub
n)$-tiling with the bars $R_{h \times 1}$ and $R_{1 \times v}$
then, by simple considering the area covered by the tiling, it is
direct to notice that $ab = amv + bnh$.

Conversely, if $ab = amv + bnh$ then, by Lemma~\ref{lemm:algebra},
there exist $p,q,a',b'\geq 1$ such that:
\begin{itemize}
\item $(p+q)=\gcd(a,b)$ and $a=(p+q)a'$ and $b=(p+q)b'$.
\item $nh=pa'$ and $mv=qb'$. 
\end{itemize}
As it appears in Figure~\ref{fig:toro-unif}, the torus $T_{a \times
b}$ can be partitioned into $(p+q)^{2}$ rectangles ${\Theta}_{i,j}$
defined for each $i,j \in \{0,\cdots,p+q-1\}$ as follows:

\begin{center}
	${\Theta}_{i,j} =$ the rectangle $R_{a' \times b'}$ whose
	upper left corner is the cell $(a'i, b'j)$.
\end{center}

Let us define for each $i \in \{0,\cdots,p+q-1\}$, the following
rectangular regions of $T_{a \times b}$:
\begin{itemize}
\item 
	${\mathcal{H}}_{i} = \bigcup_{k=i}^{i+p-1}{\Theta}_{k,i}$,
	which is simply the rectangle $R_{nh \times b'}$ whose upper
	left corner is the cell $(a'i,b'i)$,
\item 
	${\mathcal{V}}_{i} = \bigcup_{k=i+1}^{i+q}{\Theta}_{i,k}$,
	which is simply the rectangle $R_{a' \times mv}$ whose upper
	left corner is the cell $(a'i,b'(i+1))$.
\end{itemize}
It is easy to notice that every ${\Theta}_{i,j}$ belongs to exactly
one of the rectangles $\{{\mathcal{H}}_{i},{\mathcal{V}}_{i}\}_{0
\leq i < p+q}$ and that therefore the latter is a partition of the
torus $T_{a \times b}$.

In order to conclude, notice that each ${\mathcal{H}}_{i}$ is
tillable by using only horizontal bars $R_{h \times 1}$ with each
row having $n$ bars.  In the same way, each ${\mathcal{V}}_{j}$ is
tillable by using only vertical bars $R_{1 \times v}$ with each
column having $m$ bars.
\end{proof}

\begin{corollary}
If a torus $T_{a \times b}$ admits a tiling with uniform tomographic
constraints then $\gcd(a,b) > 1$.
\end{corollary}

\begin{proof}
From Theorem~\ref{prop:unif-torus} together with
Lemma~\ref{lemm:algebra}, $\gcd(a,b)=p+q$ with $p,q\geq 1$.
\end{proof}

\subsection{The rectangle case}

\begin{theorem}				\label{prop:unif-rect}
Let $a,b,h,v,m,n \geq 1$.  Let $(\sub m,\sub n) = (m\cdots m,n\cdots
n) \in \N^{a} \times \N^{b}$.  The rectangle $R_{a \times b}$ is
$(\sub m,\sub n)$-tillable with the bars $R_{h \times 1}$ and $R_{1
\times v}$ if and only if $ab = amv + bnh$, $h|a$, and $v|b$.
\end{theorem}

\begin{proof}
Let us suppose that the rectangle $R_{a \times b}$ admits an $(\sub
m,\sub n)$-tiling with the bars $R_{h \times 1}$ and $R_{1 \times v}$.
By simple area considerations, it holds that $ab = amv+bnh$.  The fact
$h | a$ follows from this observation.  Since in every column we have
$mv$ cells tiled by vertical bars, the remaining $k=b-mv$ cells are
tiled with horizontal bars.  Therefore $k$ horizontal bars are between
column $1$ and column $h$, another $k$ bars between columns $h+1$ and
$2h$, and so on.  In the same way we conclude $v|b$.

For the converse, let $ab = amv + bnh$, $h|a$, and $v|b$.  We reduce
this case to a \textsc{01-Matrix Reconstruction Problem}.  Let $p,q$
be such that $a=ph$ and $b=qv$.  Now $R_{a \times b}$ may be
partitioned into $pq$ rectangles $R_{h \times v}$ and each of these
rectangles may be tiled by using one type of bars (vertical or
horizontal).  We define a $p \times q$ 01-matrix, where each entry
corresponds to a $R_{h \times v}$ rectangle, and contains ``1'' if the
later is tiled with vertical bars, and ``0'' otherwise.  The problem
is reduced to the following: given $p,q,m,n \geq 1$ such that
$pq=pm+qn$, construct a 0-1 matrix of size $p \times q$ in such a way
that each column has $m$ 1's and each row has $n$ 0's.  The solution
is trivial. In fact, it suffices to consider the 1's as vertical bars
of unitary length (simple squares), the 0's as horizontal bars of
unitary length (simple squares), and to apply
Theorem~\ref{prop:unif-torus} (for unitary length bars the torus is
equivalent to the rectangle).
\end{proof}

\section{Horizontal bars of unit length}

In this section, we give a polynomial time algorithm for
reconstructing a rectangle tiling with horizontal bars of unit length.
We assume in this section that $h=1$.  For technical reasons we will
even give a more general algorithm for reconstructing tilings of
\emph{histograms}.

\begin{definition}
A histogram $H$ of a rectangle $R_{a \times b}$ is a subset of $R_{a
\times b}$ such that if cell $(i, j)$ is an element of $H$ with $j<b-1$,
then $(i,j+1)$ is also an element of $H$.  The \emph{top} of column
$i$ is the cell $(i,j)\in H$ with minimal $j$ (remember rows are
numbered from top to bottom).  The number of cells of the row $j$ of
$H$ is denoted by $c_{j}$.  The height of column $i$ of $H$ is the
number of cells in it.
\end{definition}

We will give an algorithm which, given a vector $ (m_{0},m_{1},\dots
,m_{a-1},n_{0},n_{1},\dots ,n_{b-1}) $ of integer coordinates, an
integer $v$ and a histogram $ H $ of rectangle $ R_{a\times b}$,
constructs a tiling of $ H $ with the bars $R_{1\times1}$ and
$R_{1\times v}$ satisfying the tomographic constraints $(\sub m, \sub
n)$ or answers ``No'' if there is no such tiling.

\subsection{Algorithm}

This algorithm is based on a very simple idea: A solution is
constructed iteratively row by row, where vertical tiles are placed in
columns of largest remaining constraint.  See
\textsl{http://www.lri.fr/\-\~{}durr/\-VertOnly/\-vertOnly.html} for
an implementation.

\begin{tabbing}
\textbf{input:} 
	$\sub m\in\N^a, \sub n\in\N^b, v>0$, histogram $H\subseteq
	R_{a\times b}.$
\\
\textbf{promise:}
	$\sum_{i=0}^{a-1} v m_i + \sum_{j=0}^{b-1} n_j = |H|$.
\\
\textbf{For} \= $j^*$ from $0$ to $b-1$ \textbf{do}
\\ 
\>\textbf{If} \= $c_{j^*}<n_{j^*}$ answer ``No'' and stop.\\
\>\textbf{While} \= $c_{j^*}>n_{j^*}$ do\\
\>\>\textbf{If} \= $j^*+v-1\geq b-1$ answer ``No'' and stop.\\
\>\>	Let $i$ be a column with maximal $m_i$ which
	satisfies $(i,j^*)\in H$.
\\
\>\>	Place a vertical bar between $(i,j^*)$ and $(i,j^*+v-1)$.
\\
\>\>	\textbf{For} \= $k$ from $j^*$ to $j^*+v-1$ \textbf{do}
\\
\>\>\>  Remove cell $(i,k)$ from $H$.
\\
\>\>\>	Update $c_{k} := c_{k}-1$.
\\
\>\>	Update $m_i:=m_i-1$.	\\
\>\textbf{While} $n_{j^*}>0$ do\\
\>\>	Let $i$ be any column which
	satisfies $(i,j^*)\in H$.
\\
\>\>	Place a horizontal bar (cell actually) on $(i,j^*)$.
\\
\>\>	Remove this cell from $H$.  
\\
\>\>	Update $n_{j^*}:=n_{j^*}-1$.
\end{tabbing}

\subsection{Analysis}

The correctness of the algorithm is a consequence of the lemma below:

\begin{lemma}
Let $T$ be a tiling of an histogram $H$ satisfying the constraints
given by vector $(m_0, \ldots, m_{a-1}, n_0, \ldots, n_{b-1})$. Let
$S_{algo}$ be the set of columns of maximal height where vertical bars
are placed in the first step of the algorithm.

Assume that there exists a solution. For each tiling $T$ which solves
our problem, let $S_{T}$ be the set of columns with maximal height in
$H$ whose top is covered by a vertical bar.  Let $T^*$ be such a
solution such that $S_{algo} \cap S_{T^*}$ is maximal. Then $S_{algo}=
S_{T^*}$
\end{lemma}

\begin{proofUsing}{by contradiction}
Since $|S_{algo}|=|S_{T^*}|$ there is a column $i_{1}$ of $S_{algo}$
which is not an element of $S_{T^*}$, and a column $i_{2}$ of $S_{T^*}$
which is not an element of $S_{algo}$.  
Fix such a pair of columns $i_1$, $i_2$.
Notice that $m_{i_{2}} \leq m_{i_{1}}$.

\begin{figwindow}[6,r,\epsfig{file=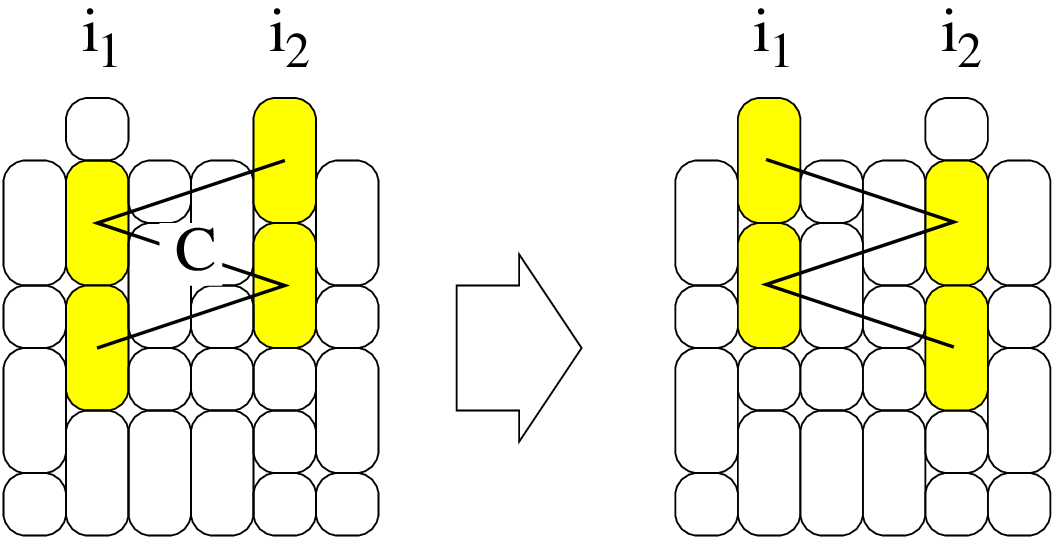,width=55mm},%
{The exchange of $I_C\times\{i_1\}$ with $I_C\times \{i_2\}$.}]
Let $V_{1}$ (respectively $V_{2}$) denote the set of vertical bars of
$T^*$ included in column $i_{1}$ (respectively column $i_{2}$). We
construct a bipartite undirected graph $G$ whose set of vertices is
$V_{1} \cup V_{2}$. Two tiles are joined by an edge if they cross a
same row.  Notice that, in $G$, each bar has at most two neighbors,
and $G$ has no cycle. Hence, $G$ is formed from disjoint chains (each
bar with no neighbor is considered as a chain of null length).  For
every chain $C$ let $I_C$ be the set of rows traversing the bars in
$C$.  Then clearly $I_C$ is an interval, and different chains have
disjoint row sets.

From such a chain $C$, one can construct a tiling $T_{C}$ transforming
$T^*$, by an \emph{exchange on chains}, which is the exchange of
$I_C\times\{i_1\}$ with $I_C\times \{i_2\}$.  Notice that this
operation preserves the tomographic constraints on the rows, while
preserving those on the columns if and only if $C$ has an even number
of vertices.

Let $C_0$ be the chain with lowest row indices. 

If $C_{0}$ has an even number of vertices, then the tiling $T_{C_{0}}$
contradicts the maximality of the intersection $S_{algo} \cap
S_{T^*}$, which, consequently, achieves the proof.

If $C_{0}$ has an odd number of vertices (i.e.{} both endpoints of
$C_{0}$ are bars in column $i_2$), then by the inequality $m_{i_2}\leq
m_{i_1}$ there exists another chain $C_{1}$ with an odd number of
vertices, whose extremities are bars in column $i_1$.  Let $T'$ be the
tiling obtained from $T^*$ by exchanging $C_0$ and $C_1$.  $T'$
satisfies the same vertical and horizontal constraints as $T^*$, and,
consequently contradicts the maximality of the intersection $S_{algo}
\cap S_{T^*}$.  This last fact achieves the proof.
\end{figwindow}
\end{proofUsing}

\begin{lemma}					\label{thm:vertOnly}
The algorithm presented in this section gives a tiling satisfying the
constraints, if such a tiling exists.  It's running time is $O(a\log a
+ab)$.
\end{lemma}

\begin{proof}
We prove its correctness by induction on the number of cells of the
histogram $H$ given as input. If $H$ is empty, the result is obvious.

Now assume that the theorem holds for each histogram which has less
cells than $H$. If $H$ admits a tiling with constraints, then, by the
previous lemma, there exists such a tiling using tiles placed in the
first execution of the loop.

After the first execution of the loop (and updating), we have to prove
the theorem for an histogram which has less cells than $H$, which is
true by induction hypothesis.

Now we turn to the proof of the time complexity.  The algorithm will
maintain an ordering on $\sub n$.
\begin{itemize}
\item 
	the initialization costs $O(a\log a)$ time units (because
	of the ordering of the columns),
\item 
	each passage through the loop costs $ O(a) $ time units,
	since the update of the order can easily be done in $ O(a) $
	time units,
\item 
	there are $b$ passages through the loop.
\end{itemize}
This proves that the execution of this algorithm costs $ O(a\log a+ab) $
time units.
\end{proof}

\begin{figure}[bt]
\centerline{\epsfig{file=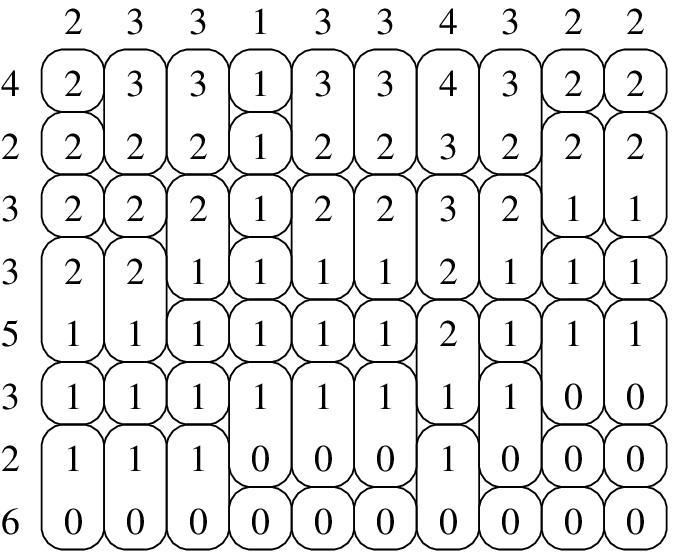,height=5cm}}
\caption{Trace of the reconstructing algorithm on
Example~\ref{ex:algo01}.}  \label{fig:algo01}
\end{figure}
\begin{example}					\label{ex:algo01} 
For the rectangle $R_{10 \times 8}$, for vertical bars of size 2 and
the projections
\[
		{\sub m} = (2,3,3,1,3,3,4,3,2,2) \in \N^{10}
\mbox{ and } 	{\sub n} = (4,2,3,3,4,3,2,6) \in \N^{8}.
\]
Figure~\ref{fig:algo01} shows the trace of the algorithm.  Numbers in
the cells indicate the remaining column constraints.
\end{example}

\subsection{Application}			\label{sec-application}

The previous algorithm can be used to reconstruct a tiling, when some
particular promise on $\sub m$ is given.  This promise is fulfilled
in particular if $\sub m$ is uniform, or monotone ($m_0\leq \ldots
\leq m_{a-1}$), as shown in~\cite{Pic01}.

\begin{theorem}				\label{prop:algo01}
Let $a,b,h,v \geq 1$.  Let $({\sub m},{\sub n}) = (m_0 \cdots
m_{a-1},n_0 \cdots n_{b-1}) \in \N^{a} \times \N^{b}$ with $m_i=m_j$
for all $i,j\in\{0,\ldots,a-1\}$ satisfying $\lfloor i/h \rfloor
=\lfloor j/h \rfloor$.  There is an algorithm in $O(a\log a+ab)$ that
decides whether the rectangle $R_{a \times b}$ is $(\sub m,\sub
n)$-tillable with the bars $R_{h \times 1}$ and $R_{1 \times v}$ and
if \emph{yes} outputs a valid tiling.
\end{theorem}

\begin{figure}[htb]
\centerline{\epsfig{file=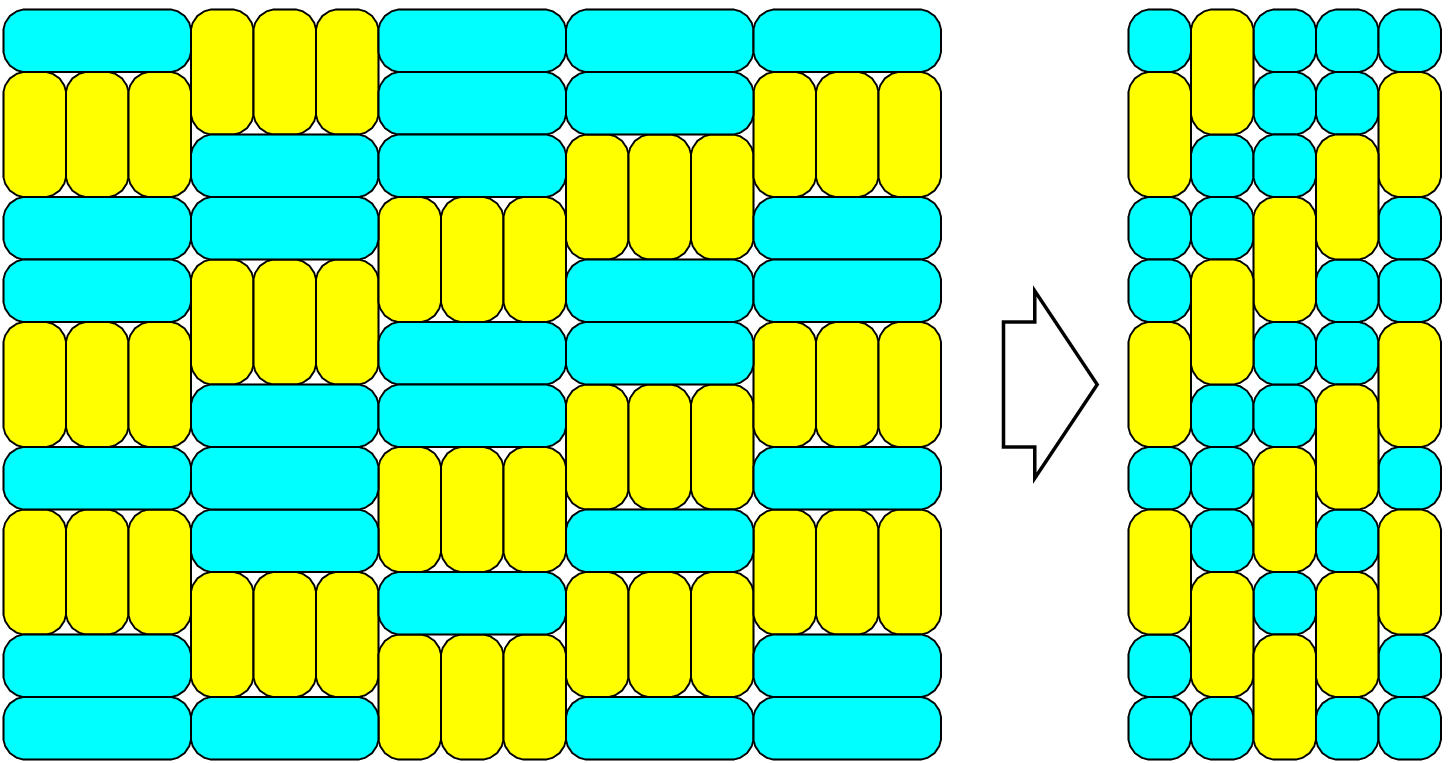,height=4cm}}
\caption{Reducing the problem by changing its horizontal scale.}
						\label{fig:scale}
\end{figure}

\begin{proof}
By the same argument of the first part of
Theorem~\ref{prop:unif-rect} it can be concluded that the tiling
of $R_{a \times b}$ may be partitioned into $\frac{a}{h}$ tilings of
rectangles of type $R_{h \times b}$.  It suffices now to divide every
horizontal measure by $h$ (i.e, to change the horizontal scale) in
order to reduce the original problem to a new one in which
$a'=\frac{a}{h}$, $b'=b$, $h'=\frac{h}{h} = 1$, $v'=v$, $(\sub m)' =
(m_0,m_h,\ldots, m_{(a'-1)h}) \in \N^{a'}$, $(\sub n)' = \sub n \in
\N^{b}$ (see Figure~\ref{fig:scale}).  We can apply now
Lemma~\ref{thm:vertOnly}.
\end{proof}

\section{Tiling a sub-grid}

In the previous section we showed that some instances of the
\textsc{Tiling with Bars Reconstruction} problem have a polynomial
solution. In this section we show that a more general problem is
NP-hard. In the \textsc{Sub-grid Domino Tiling Reconstruction Problem
From Projections} we are given a only sub-grid $ S\subseteq
R_{a\times b}$ to tile. We show that this problem is NP-hard
by a reduction from the following problem.

\begin{quote}
\noindent \textsc{The 3-Color Consistency Problem}\\
 We fix a set of colors $ \Delta =\{colorless,red,blue,green\}
 $.
\begin{description}
\item [input]
	$ \sub {m}^{c}\in \N ^{a} $ and $ \sub {n}^{c}\in \N ^{b} $
	for every $ c\in \Delta $.
\item [decide]
	if there is a matrix $ T\in \Delta ^{a\times b} $ with
	projections $ (\sub {m}^{c},\sub {n}^{c})_{c\in \Delta } $
	that is for all colors $ c $ we have
\[
			m_{i}^{c}=|\{j:T_{ij}=c\}|
	\textrm{ and }	n_{j}^{c}=|\{i:T_{ij}=c\}|.
\]
\end{description}
\end{quote}
It has been shown in \cite{CD01} that this problem is NP-hard in the
strong sense, while the \textsc{1-Color Consistency Problem} is
solvable in linear time~\cite{Ryser63}. The \textsc{2-Color
Consistency Problem} is still open.

\subsection{Gadget}

A sub-grid is a \emph{cycle} if every cell has exactly two (horizontal
or vertical) adjacent neighbors, and if every pair of cells is
connected by transitivity. We start by giving some facts about cycles.

\paragraph*{The cycle length is always even.}

This can be easily seen by coloring the cells checkerboard wise black
and white.  Then adjacent cells have different colors. The claim
follows from the fact that the cycle is closed.

\paragraph*{There are exactly two domino tilings of a cycle.}

Fix any numbering of the cells such that every cell $ i $ has
neighbors $ i-1 $ and $ i+1 $ modulo the length of the cycle. Then
clearly one tiling covers all pair of cells $ (2i,2i+1) $ with a
domino, while the other one covers $ (2i,2i-1) $ for all $i$. 

\begin{figure}[htb]
\centerline{\epsfig{file=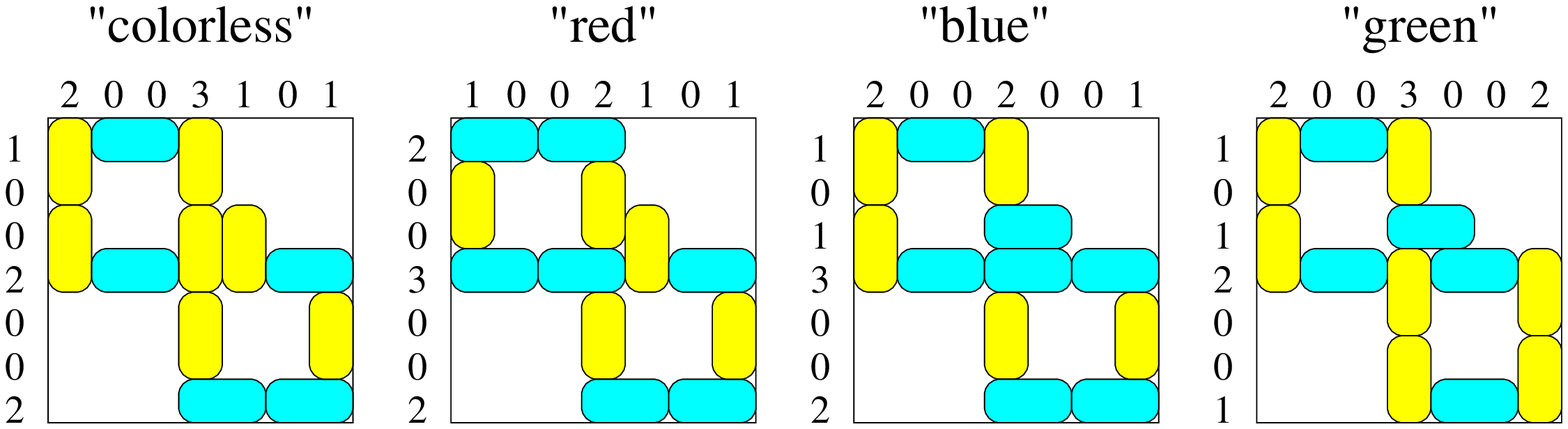,height=3cm}}
\caption{All four tilings of \protect\protect$ S\protect \protect $.}
						\label{all4}
\end{figure}

We specify now a sub-grid $ S $ consisting of two cycles intersecting
at a corner and an addition cell.  This additional cell must, in a
domino tiling, be matched to a cell of one of the cycles, therefore
``forcing'' it to admit a unique tiling, while the other one admits
the usual two tilings.  As a result we will have exactly four tilings
of $S$.  We define $S$ to be the subgrid shown with its tilings in
figure~\ref{all4}.  We refer to these tilings as $ T_{colorless},
T_{red}, T_{blue}, T_{green} $ respectively. Let $ (\sub {s}^{c},\sub
{t}^{c}) $ be their projection vectors for every color $ c $.  Note
that by the symmetry of $S$ we have $\{\sub
{s}^{c}\}_{c\in\Delta}=\{\sub {t}^{c}\}_{c\in\Delta}$.

\begin{lemma}					\label{lem:vec_{i}nd}
The vectors $\{\sub{s}^{c}\}_{c\in\Delta}$ are linear independent.
\end{lemma} \begin{proof}
Let $\sub{u}\in \N^{7} $ be an arbitrary linear composition of the
column vectors. We have to show that the coefficients in
$\sub{u}=\sum_{c}
\alpha_{c}\sub{s}^{c} $ are uniquely defined. We have
\begin{eqnarray*}
u_{1} & = & 2\alpha _{colorless}+1\alpha _{red}+2\alpha_{blue}+2\alpha_{green}\\
u_{3} & = & 3\alpha _{colorless}+2\alpha _{red}+2\alpha_{blue}+3\alpha_{green}\\
u_{4} & = & 1\alpha _{colorless}+1\alpha _{red}+0\alpha_{blue}+0\alpha_{green}\\
u_{7} & = & 1\alpha _{colorless}+1\alpha _{red}+1\alpha_{blue}+2\alpha_{green}\\
\end{eqnarray*}
This system of equations has a unique solution which concludes the
proof: $\alpha_{colorless} = -2, \alpha_{red}=3, \alpha_{blue}=2,
\alpha_{green}=-1$.
\end{proof}

\subsection{The proof of NP-hardness}

\begin{figure}[bht]
\centerline{\epsfig{file=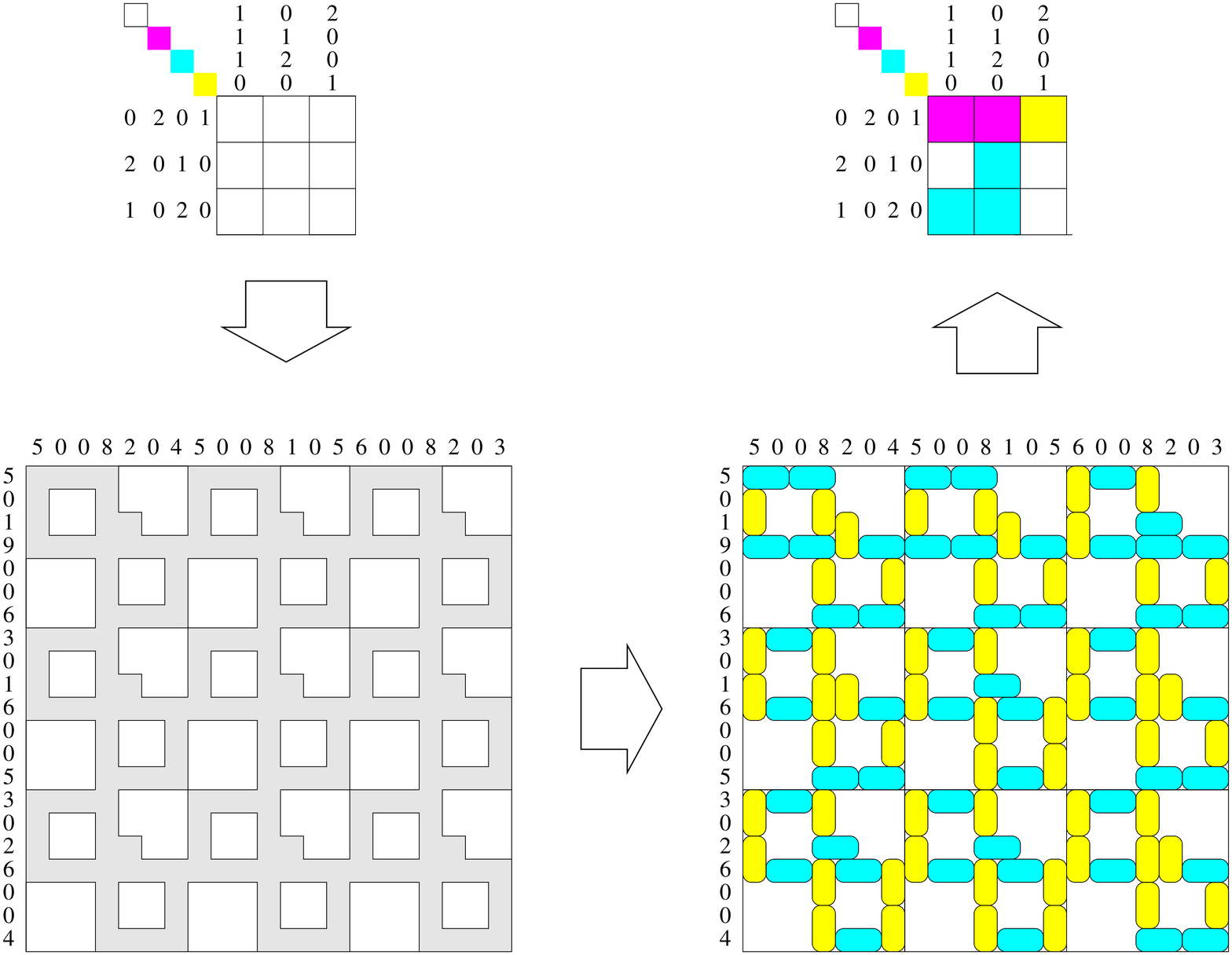,width=15cm}}
\caption{Idea of the reduction}
\label{reduction}
\end{figure}

\begin{theorem}
The \textsc{Domino Sub-grid Tiling Reconstruction Problem From
Projections} is NP-hard in the strong sense.
\end{theorem} 
\begin{proof}
Let $ I=(\sub{m}^{c},\sub{n}^{c}) $ an arbitrary instance of the
\textsc{3-Color Consistency} problem for an $ a\times b $ matrix. We
construct an instance of the former problem, such that there is a
bijection between the respective sets of solutions. This proves then
the theorem. (see figure~\ref{reduction})

We define the instance $I'$ as $(\sub{m},\sub{n})\in
\N^{7a}\times \N^{7b}$ with
\[
	\sub{m} = \bigotimes ^{a}_{i=1}\left(
	\sum _{c}m_{i}^{c}\sub{s}^{c}\right) 
		\textrm{ and }
	\sub{n} = \bigotimes ^{b}_{j=1}\left( 
	\sum _{c}n_{j}^{c}\sub{t}^{c}\right) ,
\]
where $\otimes$ denotes the concatenation of vectors and
\[
	S'=\bigcup _{i=0}^{a-1}\bigcup _{j=0}^{b-1}(S+(7i,7j)).
\]

In a tiling of $S'$ every $7\times 7$ block contains one of the four
tilings of $S$.  Therefore there is a natural bijection $f$ between
the set of domino tilings of $S$ and the set of matrices
$\Delta^{a\times b}$.  It follows from lemma~\ref{lem:vec_{i}nd} that
$T$ is a solution to the instance $I'$ if and only if $f(T)$ is a
solution of $I$.  Moreover for the unary encoding the size of $f(T)$
is linear in the size of $T$.
\end{proof}

\section{Concluding remarks}

We will conclude with a observation for the general problem.  Let
$(\sub m, \sub n, h,v)$ be an instance for the reconstruction problem
for tilings of an $a\times b$-rectangle.

We say that a particular realization is \emph{separated} between
column $i$ and column $i+1$, if there is no horizontal bar traversing
the border in between.  We claim that if this is the case for one
realization, it holds for all other realizations as well: Let $c_i$ be
the number of horizontal bars beginning in column $i$ and ending in
column $i+h-1$.  Let us also denote $c_i = 0$ for all $i<0$. Then
clearly the following induction holds:
\[
	c_i = b - (vm_i + c_{i-h+1}+\ldots +c_{i-1}).
\]
Therefore there is a separation between column $i$ and column $i+1$ if
and only if $c_{i-h+2}+\ldots + c_{i}=0$, which is a realization
independent condition.

\begin{figure}[bt]
\centerline{\tiny\begin{tabular}{c|ccc|ccc|}
	&2&1&2&2&1&2 	\\ \hline
1&&&&&&\\
0&&&&&&\\
2&&&&&&\\\hline
2&&&&&&\\
1&&&&&&\\
2&&&&&&\\\hline
	    \end{tabular}
\hspace{2cm}
\begin{tabular}{c|cccccc|c|}
	&2&1&2&2&1&2&4 	\\ \hline
3&&&&&&&\\
3&&&&&&&\\\hline
1&&&&&&&\\
0&&&&&&&\\
2&&&&&&&\\
2&&&&&&&\\
1&&&&&&&\\
2&&&&&&&\\\hline
	    \end{tabular}
}
\caption{Separations for $h=2, v=2$.}  \label{fig:cond}
\end{figure}

In the same manner we define the separation between lines.  These
separations, which can be computed in linear time, partition the grid
into \emph{separated rectangles} which are surrounded either by a
separation or by the border of the grid.  Clearly if there is a
realization of $(\sub m, \sub n, h,v)$ then
\begin{enumerate}
\item
	$(v \sub m, \sub n, h, 1)$ and $(\sub m, h\sub n, 1, v)$ must
	have a realization as well 
\item
	and every separated rectangle must admit a tiling with
	horizontal bars of length $h$ and vertical bars of height $v$,
	even without any tomographic constraint.
\end{enumerate}
Left part of figure~\ref{fig:cond} shows an instance which satisfies
the first but not the second condition, since each of the four
separated rectangle has odd size.  However the two conditions are not
sufficient: The right hand instance satisfies conditions 1 and 2.  But
it has no solution, since the last column must be filled with vertical
dominoes, the remaining cells of the first row must be tiled with
horizontal dominoes, and for the remaining rectangle we end up with
the left hand side instance.  However the two conditions are not
sufficient, as shows the right part.

\end{document}